\documentclass[preprint,showkeys]{revtex4-1}
\usepackage{graphicx}
\usepackage{amssymb}
\usepackage{amsmath}

\begin{document}

\title{Democracy and polarization in the National Assembly of the Republic of
Korea}


\author{Jonghoon \surname{Kim}}
\author{Seung Ki \surname{Baek}}
\email[]{seungki@pknu.ac.kr}
\thanks{Fax: +82-51-629-5549}
\affiliation{Department of Physics, Pukyong National University, Busan 48513,
Korea}


%
%
%
%
%


\begin{abstract}
The median-voter hypothesis predicts convergence of party platforms across a
one-dimensional political spectrum during majoritarian elections. Assuming that
the convergence is reflected in legislative activity, we study the time
evolution of political polarization in the National Assembly of the Republic of
Korea for the past 70 years. By projecting the correlation of lawmakers onto the
first principal axis, we observe a high degree of polarization from the early
1960’s to the late 1980’s before democratization. As predicted by the
hypothesis, it showed a sharp decrease when party politics were revived in 1987.
Since then, the political landscape has become more and more multi-dimensional
under the action of party politics, which invalidates the assumption behind the
hypothesis. For comparison, we also analyze co-sponsorship in the United States
House of Representatives from 1979 to 2020, whose correlation matrix has been
constantly high-dimensional throughout the observation period. Our analysis
suggests a pattern of polarization evolving with democratic development, from
which we can argue the power and the limitation of the median-voter hypothesis
as an explanation of real politics.
\end{abstract}

\keywords{Principal-component analysis, Co-sponsorship, Median-voter hypothesis,
Political polarization}



\maketitle

\section{Introduction}
Democracy is noisy. Through party politics, it invites the public into
conflicts that were once private so that the conflicts are
socialized as public
affairs~\cite{schattschneider1975semisovereign,adamany1972political}.
A conflict of great social importance becomes a cleavage around which parties
are positioned, and the parties compete for votes by representing different
views on the cleavage. Therefore, polarization of parties is an inherent
and inevitable feature of democracy, despite growing concern about its
negative consequences~\cite{layman2006party,campbell2018polarized}. At the same
time, the median-voter hypothesis (MVH) argues that two parties will converge to
the same position across a one-dimensional political spectrum as long as voters
always vote with single-peaked
preferences~\cite{downs1957economic,holcombe1989median}: To be more specific,
let us
imagine a number of voters, each of whom has his or her own political position
as a real number and votes for a party whose position is the closest to his or
her own. A party has to obtain more votes than the other parties to win the
election, and if two parties play this zero-sum game, the Nash equilibrium
is reached when both converge to the median voter's position. A rough sketch of
party politics would be an interplay between these two forces of polarization
and convergence.

The empirical evidence for the MVH is mixed, and the main reason is that
the political privacy in the secret-ballot system poses an obstacle to
its direct verification. One way to get around the difficulty is to assume that
the median voter is effectively equal to a voter with the median income (or the
median tax
price)~\cite{barr1966elementary,mathis1986examination,turnbull1994median,milanovic2000median}.
If this is the case, the MVH provides testable predictions: For example,
a more unequal society is predicted to choose greater redistribution because the
median voter will be relatively poorer in a more unequal society when compared
with the mean income~\cite{milanovic2000median}.
As an alternative indirect measure, one can compare the dispersion in
expenditures of local governments: An early finding was that local governments
with a single party showed a greater coefficient of variation than those with
two parties, suggesting that moderate policies tend to be adopted as a result of
political competition~\cite{kasper1971political,mceachern1978collective}. Still,
whether the MVH outperforms the existing models
mostly based on mean incomes remains in doubt,
and a number of studies present statistical results against the
MVH~\cite{romer1979elusive,boyne1987median,mueller2003public}.

In this work, we wish to examine the dynamics of party politics by looking at
the co-sponsorship in the National Assembly of Korea for the past 70 years.
Co-sponsorship has been used to analyze collaboration patterns among
lawmakers~\cite{canen2021endogenous}:
In the United States, co-sponsorship is known to exhibit
a lower degree of polarization than roll-call voting because successful
legislation requires a supermajority in the presence of bicameralism, the
filibuster, and the presidential
veto~\cite{harbridge2015bipartisanship,curry2019non}.
However,
this work is relatively free from such a risk of underestimation because the
immediate goal of co-sponsorship in Korea is not to pass the bill but to
introduce it officially to the Assembly: It needs only 10 lawmakers, and the
bill usually expresses its own idea most clearly at this stage before undergoing
revisions to seek a compromise. Indeed, co-sponsorship analyses of the National
Assembly of Korea have detected strong
partisanship~\cite{park2017cosponsorship,baek2020co}.
We also stress that we are discussing how lawmakers perceive the party politics
as a whole: If we move our focus to each instance of congressional voting,
it will usually be explained by one or two
dimensions~\cite{mccarty2019polarization}.

Our main finding is that the dynamics of party politics tends to contradict the
assumption of one dimensionality behind the MVH. We suggest that a better
understanding would be to consider an inverse proposition of the MVH, i.e.,
predicting the lack of convergence in the absence of party politics.
In addition, we propose a scenario of how political polarization would evolve in
the course of democratic development.
This work is organized as follows: The next section discusses our methods,
including data curation and statistical observables.
We present the result in Section~\ref{sec:result} and discuss its implications
in Section~\ref{sec:discussion}. We then conclude this work in
Section~\ref{sec:summary}.

\section{Method}

\subsection{Data}

We have collected the bill data from 1950 to 2019 through
Open Data Portal (\url{http://www.data.go.kr}).
For each bill, we have its date of motion, title, chief author, and
co-sponsors. The first National Assembly (May 31 1948 -- May
30 1950) has been excluded from our analysis because it was a constituent
assembly with no co-sponsored bills. We also exclude the years 1962, 1971, 1979,
and 1992 when no or few bills were sponsored. See Table~\ref{table:data} for
details of the data.

To quantify the level of democracy, we use the {\it Freedom in the World}
reports from 1973 to 2020 by Freedom House
(\url{https://freedomhouse.org}).
Those reports have two indices on a scale from 1 (most free) and 7 (least free):
One measures political rights, and the other addresses civil liberties.

For comparison, we have also collected co-sponsorship data in the United States
House of Representatives from \url{http://www.congress.gov}. The period ranges
from 1979 to 2020.

\begin{table}
\caption{Summary of data used in this work.
The first column means the session number of the Assembly, and the second and
the third columns show the dates of establishment and disbandment, respectively.
The fourth column shows the number of seats, and the figures in the parentheses
mean the numbers of lawmakers awarded through proportional representation or
through recommendation of the president (marked by $\ast$). Single-member
constituencies
comprise the other seats in most cases, but we have a few exceptions of the
medium constituency system as noted in the last column.
The fifth column shows the number of collected bills.
For the 20th Assembly, we use the bill data in Ref.~\cite{baek2020co}, which was
collected up to June 3, 2019 ($\dagger$). The Assembly has been unicameral
except for the 5th session.
}
\begin{tabular}{crrccc}
\hline
session \# & established & disbanded & \# of seats & \# of bills & misc.\\\hline
 1 & 5/31/1948 & 5/30/1950 & 200(0) & N/A & \\
 2 & 5/31/1950 & 5/30/1954 & 210(0) & 674 & \\
 3 & 5/31/1954 & 5/30/1958 & 203(0) & 625 & \\
 4 & 5/31/1958 & 7/28/1960 & 233(0) & 275 & \\
 5 & 7/29/1960 & 12/16/1963 & 291(0) & 327 & bicameral \\
 6 & 12/17/1963 & 6/30/1967 & 175(44) & 538 & \\
 7 & 7/1/1967 & 6/30/1971 & 175(44) & 238 & \\
 8 & 7/1/1971 & 10/17/1972 & 204(51) & 66 & \\
 9 & 3/12/1973 & 3/11/1979 & 219(73$^\ast$) & 221 & medium \\
10 & 3/12/1979 & 10/27/1980 & 231(77$^\ast$) & 127 & medium \\
11 & 4/11/1981 & 4/10/1985 & 276(92) & 266 & medium \\
12 & 4/11/1985 & 5/29/1988 & 276(92) & 304 & medium \\
13 & 5/30/1988 & 5/29/1992 & 299(75) & 683 & \\
14 & 5/30/1992 & 5/29/1996 & 299(62) & 427 & \\
15 & 5/30/1996 & 5/29/2000 & 299(46) & 1021 & \\
16 & 5/30/2000 & 5/29/2004 & 273(46) & 1890 & \\
17 & 5/30/2004 & 5/29/2008 & 299(56) & 6106 & \\
18 & 5/30/2008 & 5/29/2012 & 299(54) & 11564 & \\
19 & 5/30/2012 & 5/29/2016 & 300(54) & 15806 & \\
20 & 5/30/2016 & 5/29/2020 & 300(47) & 20967$^\dagger$ & \\
\hline
\end{tabular}
\label{table:data}
\end{table}

\subsection{Analysis}
To examine the data from the viewpoint of the MVH, we have used principal
component analysis (PCA).
Let us consider $N$ data points, each of which is $M$-dimensional.
For the co-sponsorship data, $N$ and $M$ are the number of lawmakers and the
number of bills, respectively, for a given period.
The whole data can then be represented by an $M \times N$ matrix $R$, and
a column vector $[R_{1n}, \ldots, R_{Mn}]^\intercal$ denotes the $n$th data
point, where $\intercal$ means transpose. We set $R_{mn}=1$ if the $n$th
lawmaker sponsored the $m$th bill, and $R_{mn}=0$ otherwise.
After removing all-zero rows and columns, we work with a standardized matrix:
\begin{equation}
X = \begin{pmatrix}
\frac{R_{11} - \mu_1}{s_1} & \ldots & \frac{R_{1N} - \mu_N}{s_N}\\
\vdots  & \ddots & \vdots\\
\frac{R_{M1}-\mu_1}{s_1} & \ldots & \frac{R_{MN}-\mu_N}{s_N}
\end{pmatrix},
\end{equation}
where $\mu_n \equiv M^{-1} \sum_m R_{mn}$ and $s_n = \sqrt{\sum_m
(R_{mn}-\mu_n)^2/(M-1)}$ are the sample mean and standard deviation of the $n$th
column vector, respectively. The next step is to construct a correlation matrix:
\begin{eqnarray}
&Q& = \frac{1}{M-1} X^\intercal X\nonumber\\
&=& \frac{1}{M-1}
\begin{pmatrix}
X_{11} & \ldots & X_{M1}\\
\vdots & \ddots & \vdots\\
X_{1N} & \vdots & X_{MN}
\end{pmatrix}
\begin{pmatrix}
X_{11} & \ldots & X_{1N}\\
\vdots & \ddots & \vdots\\
X_{M1} & \vdots & X_{MN}
\end{pmatrix},
\end{eqnarray}
whose element $Q_{ij}$ is the correlation between the $i$th and the $j$th data
points:
\begin{eqnarray}
Q_{ij} &=& \frac{1}{M-1} \sum_{m=1}^M X_{mi} X_{mj}\nonumber\\
&=& \frac{1}{M-1} \sum_{m=1}^M \left( \frac{R_{mi}-\mu_i}{s_i} \right) \left(
\frac{R_{mj}-\mu_j}{s_j} \right).
\end{eqnarray}
Note that $Q_{ij}$ takes a value from $[-1:1]$ with $Q_{ii} = 1$. If $Q_{ij}$ is
close to one, lawmakers $i$ and $j$ have highly correlated
co-sponsorship behavior. If $Q_{ij}$ is negative, on the other hand, the
lawmakers are anti-correlated, meaning that they usually do not work together.
We diagonalize the correlation matrix $Q$, e.g., by using the
singular-value decomposition, and take the principal eigenvector corresponding
to the largest eigenvalue. This gives a one-dimensional representation of
co-sponsorship among lawmakers, and the validity of this dimensionality
reduction is measured by the proportion of variance explained by the principal
component. If the Assembly is highly polarized, we will find two
clusters of lawmakers on opposite sides of the axis, and the distance
between the clusters will explain most of the variance in the data.

\subsection{Observables}

To quantify political polarization, we compute the bimodality
coefficient~\cite{freeman2013assessing,pfister2013good}:
\begin{equation}
\beta = \frac{\gamma^2+1}{\kappa},
\label{eq:bc}
\end{equation}
where $\gamma$ is the skewness and $\kappa$ is the kurtosis of the distribution.
The rationale behind this quantity is that a bimodal distribution often has
asymmetric peaks and light tails, both of which increase the value of $\beta$.
For the normal distribution, we get $\beta=1/3$ whereas the maximum value of
$\beta=1$ is obtained for the bi-delta distribution.
The uniform distribution with $\beta=5/9$ can be regarded as a borderline
between unimodal and bimodal ones.
For $N$ samples on the principal axis, the formula should be rewritten as
\begin{equation}
b = \frac{g^2 + 1}{k + \frac{3(N-1)^2}{(N-2)(N-3)}},
\label{eq:sample_bc}
\end{equation}
where $g$ is the sample skewness and $k$ is the sample excess kurtosis.

Another observable in PCA is the proportion of variance in the first
principal component defined as follows:
\begin{equation}
v = \frac{\lambda_1}{\sum_{n=1}^N \lambda_n} = \frac{\lambda_1}{N},
\label{eq:v}
\end{equation}
where $\lambda_n$ is the $n$-th eigenvalue of the correlation matrix $Q$ sorted
in descending order, i.e.,
\begin{equation}
\lambda_1 > \lambda_2 > \ldots > \lambda_N.
\label{eq:eig}
\end{equation}
Note that we have used $\sum_{n=1}^N \lambda_n = N$ because $\text{Tr}(Q) =
\sum_{n=1}^N Q_{ii} = N$ is invariant under a similarity transformation.
This quantity $v$ in Eq.~\eqref{eq:v} can be used to
estimate the effective dimensionality $d_\text{eff}$
of the given data for the following reason:
Formally speaking, $d_\text{eff}$ is commonly
estimated by selecting the smallest number of principal
axes up to which the cumulative fraction of variance exceeds a certain level
$\Lambda$ somewhere between $70 \textendash
90\%$~\cite{jolliffe1986principal,wall2002singular}.
Combining this method with the eigenvalue structure [Eq.~\eqref{eq:eig}], we
see that
\begin{equation}
\frac{\sum_{n=1}^{d_\text{eff}-1}\lambda_n}{N}
< \Lambda <
\frac{\sum_{n=1}^{d_\text{eff}}\lambda_n}{N} < \frac{\lambda_1 d_\text{eff}}{N}
= v d_\text{eff}.
\end{equation}
In this sense, we argue that $\Lambda/v$ provides a lower bound for
$d_\text{eff}$. If $v \sim O(1)$, for example, the
correlation is nearly one-dimensional, or we can say that it is
``thin'', and the first principal axis explains most of the variance in the
data. If $v \ll 1$, on the other hand, the correlation structure has high
effective dimensionality that cannot be reduced adequately along a
one-dimensional axis; hence, it is ``thick''.

For polarization to be a meaningful description of party politics, the lawmakers
need to occupy two diametrically opposite sides of the political landscape,
making every issue nearly one-dimensional. In other words, politics can be
diagnosed with polarization when both $b$ and $v$ are high in terms of the
PCA results.
More specifically, we set the criterion as $b$ greater than $5/9$, the
borderline between unimodal and bimodal distributions, in addition to $v \gtrsim
0.3$ so as to drop the lower bound of $d_\text{eff}$ down to
$O(1)$~\cite{serway2014physics}.

\section{Result}
\label{sec:result}

\begin{figure}
\includegraphics[width=\textwidth]{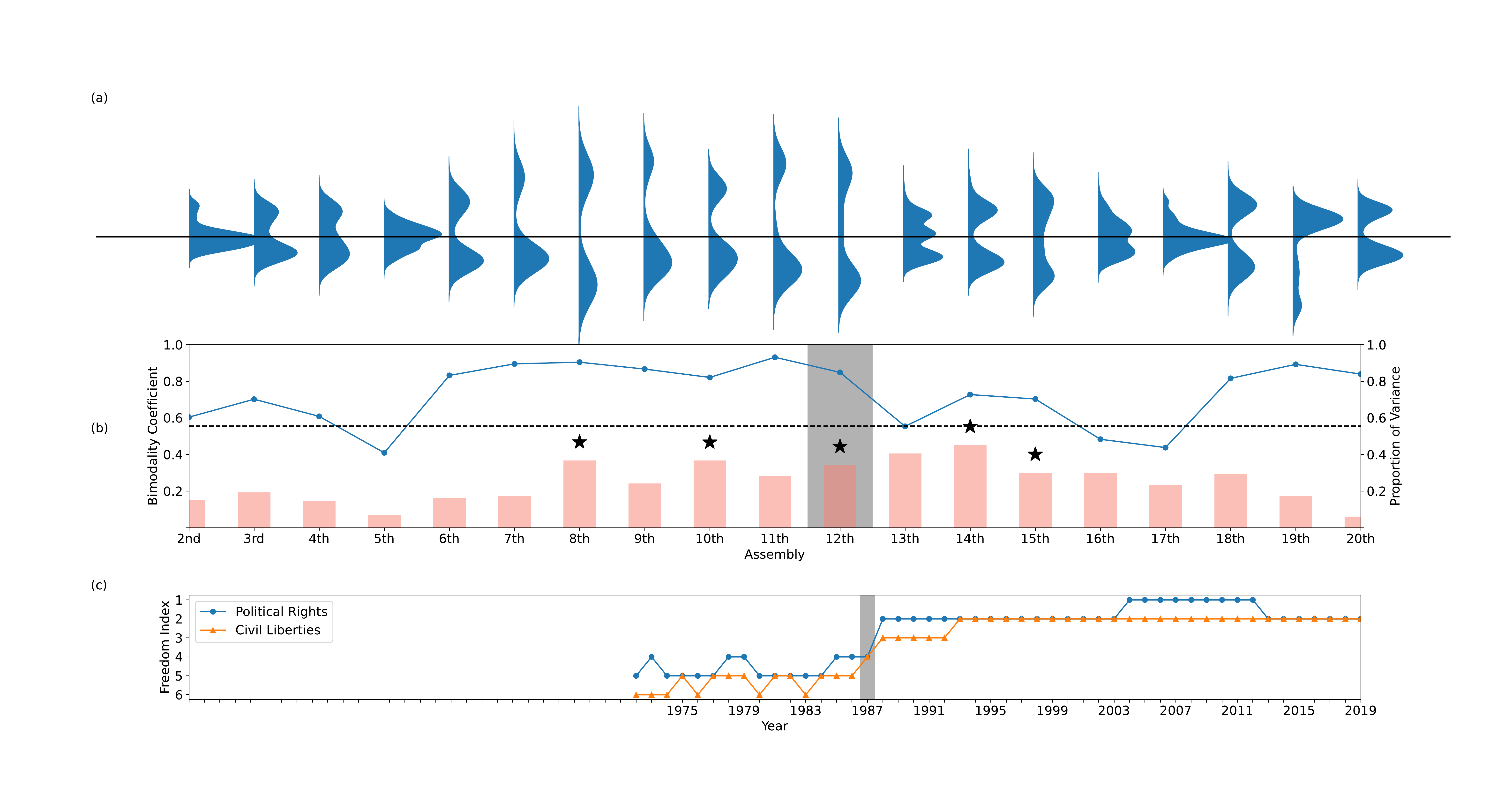}
\caption{Time evolution of political polarization in the National Assembly of
the Republic of Korea. (a)
Each year's distribution of lawmakers along the first principal axis.
When two sessions exist in the same year because of an election,
they are treated separately.
When the distribution for an Assembly is bimodal,
the clusters are almost always divided by party membership.
(b) Sample bimodality coefficients [Eq.~\eqref{eq:sample_bc}]
depicted as the line graph. The dashed horizontal line is for $\beta = 5/9$ of
the uniform distribution, which serves as a borderline between unimodal and
bimodal distributions, and the bar graph shows the proportion of variance
in the first principal component, denoted by $v$ [Eq.~\eqref{eq:v}].
We have drawn asterisks when $b>5/9$ and $v>0.3$, and
the shaded rectangle indicates the establishment of the Sixth
Republic in 1987.
(c) Political rights and civil liberties in Korea, according to
{\it Freedom in the World}. Again, we have drawn a shaded rectangle to indicate
the year 1987.
}
\label{fig:polar}
\end{figure}

With the MVH as a working hypothesis,
let us look at the political polarization in the National
Assembly of the Republic of Korea for the past 70 years.
The whole period can be divided into three parts:
The first corresponds to the First Republic, which was ended in 1960 by the
April Revolution against autocracy. Shortly after the revolution, Korea
experienced military dictatorships from 1961 to 1987. Opposition parties did
exist, but they were always minorities by the rules of the game. The last period
started in 1987 with the establishment of the Sixth Republic, after which Korea
has been on the road to democracy. We will mainly contrast the last period
(1988 -- present) with the second under military governments (1961 -- 1987)
because they are similar in length but substantially different in the political
ideas.

Our question is how
the different political structures have affected the degree of polarization,
as measured by lawmakers' correlation in their co-sponsorship of
bills
\cite{fowler2006connecting,porter2007community,zhang2008community,harward2010calculus,macon2012community,rombach2014core,andris2015rise,colliri2019analyzing,baek2020co}:
According to the MVH, it will decrease after democratization whereas
we would have no reason to expect such an effect under a dictatorship.
Our analysis indeed shows that the lack of democracy induced a high level
of polarization, which in turn shrank the political landscape to
nearly one dimension:
Figure~\ref{fig:polar}(a) shows each year's distribution of lawmakers along the
first principal axis, and the corresponding bimodality coefficient $b$
[Eq.~\eqref{eq:sample_bc}], depicted as a line graph in
Fig.~\ref{fig:polar}(b), remained high under military
governments from the early 1960s to the late 1980s.
Consistent with the MVH,
the accumulated tension was first released in the late
1980s when democracy began to work in the Sixth Republic, so
the bimodality showed a meaningful decrease in the
13th Assembly (1988 -- 1992), which was a period of transition
from military dictatorship to formal democracy [Fig.~\ref{fig:polar}(c)].
Thereafter, the line graph of
$b$ shows wild fluctuations, but the difference from the earlier period is
striking.

Another notable feature is that the proportion of variance explained by
the first principal component, denoted by $v$ (the bar graph),
had a clearly increasing tendency from the beginning of the dictatorship [see
the 5th Assembly (1960~--~1963) in Fig.~\ref{fig:polar}(b)]
and eventually went beyond $50\%$ around the transition period in the early
1990s, which implies that the effective political dimension was close to one.
Since then,
the bar graph has decreased down to $10\%$, which
means a considerable increase in the effective dimensionality. In other words,
politics has gradually become higher dimensional over the past 30 years.
We have
performed the permutation test for the time series of $v$ from the 14th
Assembly, and the resulting $p$-value is less than $1\%$.

\section{Discussion}
\label{sec:discussion}

When parties take over control over the government through elections,
taking extreme positions across the political spectrum
becomes risky for them.
On the other hand, a dictatorship would hardly feel motivated to
compromise with others' opinions:
To evade such pressure towards the median
systematically, military dictatorships abused the medium-constituency system in
the 1970s and 1980s, and some lawmakers were even awarded through recommendation
of the president (Table~\ref{table:data}). The result was a high level of
political polarization without relaxation for decades.

The establishment of the Sixth Republic in 1987 was a milestone in the political
history of Korea, which
allowed party politics to revive after a long military dictatorship.
The government party failed to win a majority in the 1988 legislative election
for the 13th Assembly, for the first time since 1950, and the government
party had to merge with two opposition parties in 1990 to regain the majority.
Since 1988, the degree of political polarization has fluctuated widely, and this
should not be surprising if we consider that parties obtain power by socializing
conflict and representing different views on it.
Exactly which factor determines the varying degrees of
polarization is unclear: For example, the bimodality in the 16th and the 17th
Assemblies is unexpectedly low although they were typically marked by
serious conflicts including a violent clash over presidential
impeachment in 2004.

A more noticeable trend is that the principal axis occupies less and less of the
total variance as time goes by. Despite the rise in the bimodality coefficient
in the last decade, its actual contribution to polarization is far
less significant than it was in the mid-1980's.
This observation implies that the current party politics in
the Assembly can no longer be described adequately along a one-dimensional axis:
It should rather be understood as a combination of multiple cleavages that are
simultaneously in action, and the MVH ceases to be a realistic description at
this stage.

\begin{figure}
\includegraphics[width=\textwidth]{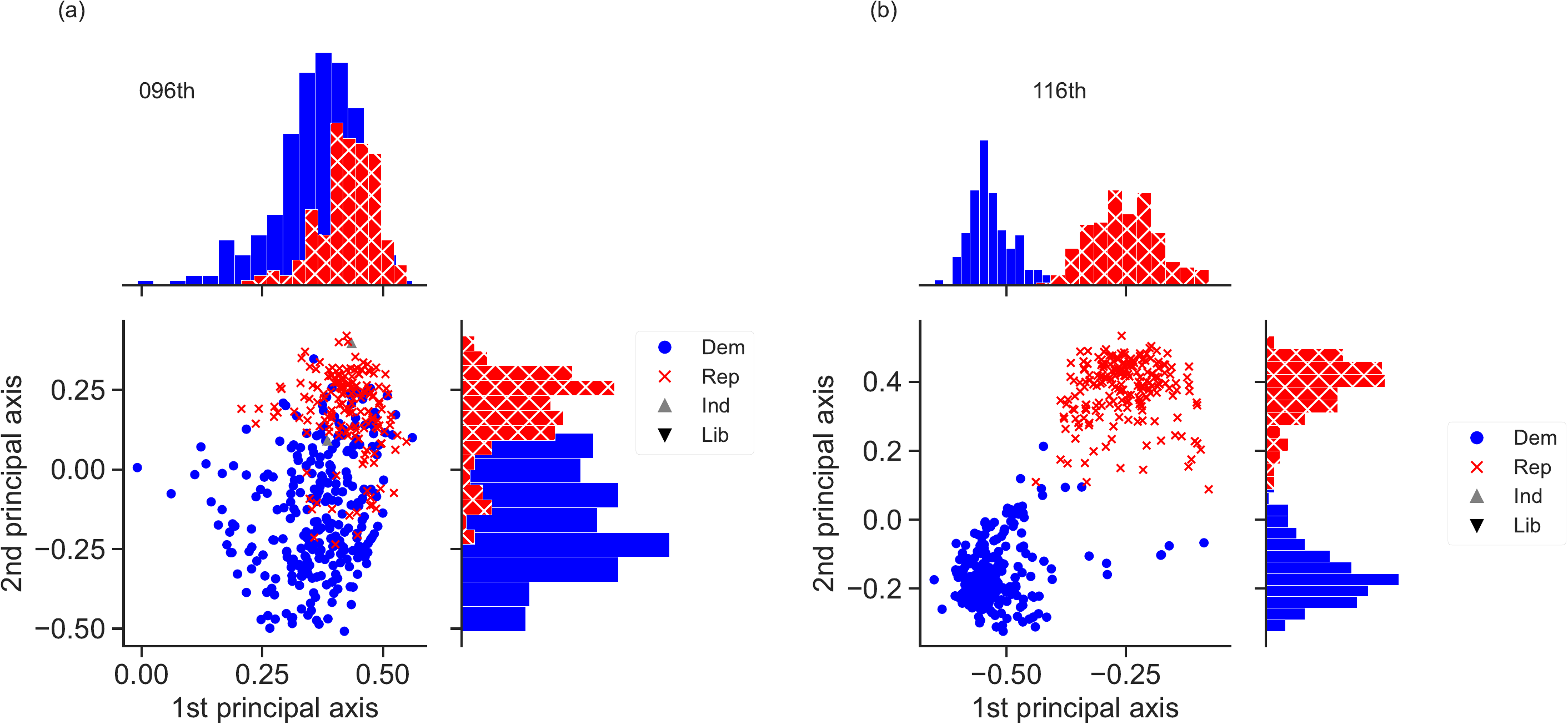}
\caption{Two-dimensional representation of co-sponsorship in the United States
House of Representatives through PCA. The blue, red, and black dots mean
Democrats, Republicans, and independent lawmakers, respectively.
(a) The 96th Congress (January 15, 1979 -- December 16, 1980) showed a
considerable overlap between Democrats and Republicans, which was actually
a common pattern before 2018.
(b) Recent segregation of Democrats and Republicans is clearly visible
in the 116th Congress (January 3, 2019 -- January 3, 2021). Still, the first
principal axis contains only 20\% of the total variance, and the second axis
does less than 10\%.
}
\label{fig:us}
\end{figure}

In this respect, our point is that the MVH could be
better supported by non-vanishing polarization in the absence of democracy,
where the political spectrum becomes nearly one-dimensional between despotism
and democracy. On the other hand, the intrinsic dynamics of party politics,
which the MVH intends to describe, will
eventually undermine its assumption on one-dimensionality because parties are,
by nature, apt to introduce new dimensions of conflict into the public sphere.
If we look at the United States House of Representatives as an example of a
mature political system (Fig.~\ref{fig:us}), even today, for which the
bimodality coefficient $b \gtrsim 0.7$ signals a record-high level of
polarization, the political space of the United States is still far from
one-dimensional in the sense that $v \approx 20\%$, which is actually a little
smaller than the $25\%$ in the mid-1980s.

\begin{table}
\caption{The three phases of polarization and democracy proposed in this work.
The first phase corresponds to the pre-democratic state of Korea before 1987.
The current constitution, the Sixth Republic of Korea, is still in the second
phase [see Fig.~\ref{fig:polar}(b)]. The third phase describes the United States
House of Representatives, especially before 2018.}
\begin{tabular}{cccc}\hline
phase & dimensionality & bimodality &  description\\ \hline
I & close to one & high & pre-democratic\\
II & increasing & fluctuating & democratization\\
III & high & low & mature democracy\\
\hline
\end{tabular}
\label{table:phase}
\end{table}

Based on these observations, we propose
three phases of polarization in the course of political development
(Table~\ref{table:phase}): In the first phase, where party politics is under the
control of an autocratic government, every conflict is projected onto a
single political axis and is usually intensified without relaxation, so the
situation is easily reduced to  ``You are either with us, or against us.''
In the second phase, the tension is released by the action of party politics.
At this stage, parties discover additional dimensions of conflict, whereby
the boundary between `with us' and `against us' gets blurred, which may lead to
frequent changes in the political alignment of parties. Finally, in the last
phase, mature party politics secures enough thickness to absorb conflicts and
maintain social integration.
Note that the one-dimensional party dynamics as assumed in the MVH holds true
only in a very narrow region between the first and the second phases.

Admittedly, such a three-phase picture is a largely speculative one, calling for
more empirical verification. Also, from the theoretical perspective,
we are not dealing with the MVH in its rigorous sense: For example, the
distinction between the median and the mean might be virtually
nonexistent in political terms such as moderates, and empirically
comparing the convergence point of lawmakers with the actual political position
of the median voter among the electorate would be difficult. Our goal was rather
to capture the nontrivial dynamics of political convergence and polarization
by looking at the co-sponsorship data from the perspective of the MVH.

\section{Summary}
\label{sec:summary}
In summary, we investigated the time evolution of political polarization in
the National Assembly of Korea from 1950 to 2019 by applying PCA to
co-sponsorship data. The degree of polarization was measured by using the
bimodality coefficient of the distribution of lawmakers along the first
principal axis, and we checked the proportion of variance explained by the first
principal
component. For the period of dictatorship from the early 1960s to the late
1980s, we observe no relaxation of political polarization, and the proportion
of variance in the first principal component accumulated over time
throughout this period. The political polarization was greatly reduced by the
establishment of formal democracy in 1987. Since then, the level of polarization
has fluctuated, but the proportion of variance in the first principal component
has constantly decreased. The additional analysis of the United States House of
Representatives suggests that both of these observables will eventually stay at
low values in a mature democracy. Our observation provides a dynamic view of
party politics in terms of the socialization of conflict, a noisy, yet
inevitable, process of democracy.

\acknowledgments
This work was supported by a research grant of Pukyong National University
(2020).

\section*{Declarations}

Not applicable.

%
\end{document}